# Breaking symmetry through speed-induced beam-deflection via centrifugation of the light source

*G. Sardin*
*Applied Physics Department, University of Barcelona - Barcelona, Spain*

**Abstract**

The experiment proposed aims to evidence and to measure the deflection of light rays induced by the source speed upon emission, and to discern it from the aberration of light rays induced by the observer speed. The method stands in creating a speed asymmetry between that of the source and that of the observer, and relies on the centrifugation of a light source at ultrahigh speed. When source and observer have the same speed, such as when being both on the same inertial system, the deflection and the aberration compensate and their net effect is null. So, in order to circumscribe the odds of inertial systems and to isolate these speed-induced effects it has instead been appealed to a centrifugal system, since it allows infringing symmetry between the source and the observer speeds. Three cases are considered. (a) Centrifugation of the source fixed at one end of the rotor arm, while the detector is fixed on the rotor axis. This configuration of the experiment aims to unveil that due to the peripheral speed of the source the beam is deflected forward, so it impinges on the centric detector slightly shifted from its position when the system was not rotating. (b) The positions of the source and the detector are interchanged. It is hence the turn of the source peripheral speed to be null and thus no speed-induced deflection ensues, but then due to the observer peripheral speed there is a speed-induced aberration, i.e. by the time the light reaches the detector this one has slightly moved side-way. However, this time the spot shifts in opposite direction and hence the effect of centrifugation is not reciprocal under the interchange between source and observer. (c) The source and the detector are fixed at the two diametrical ends of the rotor arms. In this case the beam deflection and aberration add, since their speed vectors have opposite directions and each shift of the spot on the detector is double since the time-of-flight is then double. So, in this rotating experiment the reciprocity of the effect of speed does not hold on, in the sense that if source and observer are interchanged, the corresponding effect on the light beam is not reciprocal, and has a distinct cause, i.e. deflection in one case and aberration in the other one. This lack of reciprocity differentiate them from inertial systems in which the two effects compensate since their speeds vectors have the same direction, leading thus to a null result that makes them profitless for the present aim of isolating the speed-induced deflection of light rays.

**Introduction**

We have introduced elsewhere (1-4) the speed-induced deflection of light as an actual physical effect, and it should not be mistaken with an aberration. Here, it will be distinguished between the speed-induced beam-deflection arising from the source motion and the speed-induced aberration arising from the observer motion. So, we present an experiment pursuing the specific detection of the speed-induced deflection and of the speed-induced aberration. This cannot be achieved on inertial systems using local light sources since then, the source and observer speeds being equal, the speed-induced deflection of the emitted rays due to the source motion is cancelled out by the speed-induced aberration due to the equality of the observer motion, and hence the net effect is null, i.e. the forward shift of the spot due to the source motion is compensated by the forward shift of the detector due to its motion, and hence no net spot shift ensues and the spot stay still on the detector array. To turn out this obstacle the experiment proposed has been conceived in order to create an asymmetry between the source speed and that of the observer. The light source, which can conveniently be a tiny solid state laser or diode, is rotated at an ultra-high speed, while the detector is fixed to the centrifuge rotor, so for practical means the detector peripheral velocity can be considered null (for $r \to 0 \Rightarrow v = \omega.r \to 0$). The photo-detector is constituted by a high-resolution array, which is always facing the laser since being fixed at the rotor axis.

**Experimental**

Centrifuges rotate at very high speeds, typically from 50000 to 100000 rpm (5,6). Let us pick up an intermediate angular speed: f = 60 000 rpm = 1000 cycles/s. For an arm length of 3 m the peripheral speed would be:

v = (2πr).f = 2π x 3 x 1000 ≈ 18.850 m/s ≈ 18.85 km/s

The rotating system should be put under high vacuum to ease reaching ultra-high speed and for the beam to be free of any eventual aberration. There are engineering and materials challenges in producing such an equipment. The detector should be carefully isolated from any vibration of the centrifuge. Also the huge centrifugal force (f = $mv^2/r$) due to the huge acceleration attained of million times that of gravity on earth, exerted upon the light source (diode laser) may eventually slightly alter its performances. Also, this centrifugal force acting upon the device arms would greatly stretch it, the material resistance being thus the factual most restrictive parameter.

**(1) Deflection induced by the source speed**

In this first arrangement the source is fixed at the end of the rotor arm and the detector on the rotor axis (fig.1). At first the spot position is recorded, the system being at rest. After while, it is turned on, so when the source start rotating its peripheral speed induces a forward deflection of the beam (1-4). The corresponding beam deflection would be:

tn α = v / c = 18.85 $10^3$ / $3.10^8$ ≈ 6.3 $10^{-5}$

The deflection angle α is hence equal to 63 μrad.

The corresponding shift of the beam spot on the high-resolution detector array will then be:

Δx = r. tn α = 3 x 6.3 $10^{-5}$ ≈ 19.$10^{-5}$ m = 190 μm

Since the beam is deflected in the direction of motion so does the spot on the detector array (fig.1). Inverting the rotation would invert the deflection and hence the direction of the spot shift, so the two spots separation would be twice 190 μm. Thus for an array resolution of 5 μm the total shift of 380 μm would correspond to 76 pixels, and thus it would be easily discernible.

When the light source is at one end of the gyrator arm of the centrifuge the beam deflection is only due to the speed of the source, since then the detector peripheral speed is null. Since the source is rotating at a peripheral velocity of 18.85 km/s there is an actual deflection angle of 63 μrad. Even though the source is rotating the detector is always in the same position relative to it, since it rotates at the same angular velocity and thus it always faces the source. However, by the time the beam reaches the flat detector array this one has slightly rotated of an angle β = 2π(t/T) = 2π(r/c).f = 2π($3.10^3$/$3.10^8$ = 2π.$10^{-5}$ rad. Since this angle is extremely small, the corresponding insignificant inclination will not affect the measure of the spot shift, and can be neglected.

Alternatively, the system may comprise two lasers diametrically located in order to get at once two spots on the detector. The centrifuge being turned off, the two beams are initially adjusted so that their spots superpose on a centric half-transparent screen or two oppositely sided array detectors Once the centrifuge running, the beams would deflect in opposite direction and thus the spots would get apart increasingly with speed up to its maximum. The spots shift would hence be double than for a single beam: 2 Δx = 2.(190) = 380 μm.

**(2) Aberration induced by the detector speed**

The laser is now fixed on the rotor axis and the detector at the end of its arm (fig.2). This time there is no deflection of the beam since the source peripheral speed is then null, however there is a speed-induced aberration due to the detector peripheral speed. This aberration is due to the fact that while the beam reaches the detector this one has slightly moved sideway, and thus the spot is slightly shifted from its position when the detector is at rest. This produces an apparent deflection of the beam, i.e. a speed-induced aberration.

Spot shift due to the beam time-of-flight:

$\Delta x = v \cdot t = (18.85 \cdot 10^3) \cdot (3 / 3 \cdot 10^8) = 18.85 \cdot 10^{-5} \approx 19 \cdot 10^{-5}$ m = 190 $\mu$m

This corresponding aberration will be: $\tn \alpha = v / c = 18.85 \cdot 10^3 / 3 \cdot 10^8 \approx 6.3 \cdot 10^{-5}$

The aberration angle $\alpha$ is hence equal to 63 $\mu$rad.

If the rotation is inverted so does the deflection and thus the total shift will be equal to 380 $\mu$m.

**(3) Simultaneous speed-induced deflection and aberration**

In this third configuration, the laser and the detector are both fixed at the opposite ends of the rotor arm, the beam path being then equal to the centrifuge diameter (fig.3). In this arrangement there is a speed-induced deflection of the beam to which appends the aberration due to the detector sideway shift during the beam time-of-flight. Let us point out that the spot shift due to the aberration and the shift due to the beam deflection have opposite directions, since source and detector being diametrically opposed their speed vectors have reversed directions. On an inertial system the source and observer speed vectors are equal and point at the same direction, so the speed-induced deflection and the speed-induced aberration cancel out, and lead thus to no observable effect. Instead, in this rotary system the deflection and the aberration add since at any time the source and detector speeds have counter directions (fig.3), being diametrically situated. Therefore, the spot shifts due to the deflection and to the aberration add, leading to a total shift:

$\Delta x = 2 \cdot (190) = 380$ $\mu$m

If the rotation is inverted so do the shifts and thus the total shift will be of two $\Delta x$, i.e. equal to 760 $\mu$m or equivalently to 152 pixels of 5 $\mu$m each.

**Conclusion**

The point is that this experiment allows identifying separately the speed-induced deflection due to the source motion and the speed-induced aberration due to that of the observer. This distinction is of crucial interest on conceptual grounds. This cannot be achieved with an inertial system since these two effects of speed then compensate, source and observer having the same speed and hence the net effect is null. So, since on an inertial system no effect related to its own speed is observed, it may be mistakenly deduced that speed has no effect at all, a conclusion which falsifies reality. Actually two offsetting effects are taking place, so the effect of speed is present but hidden, which is not at all the same that the formal absence of any effect. A theory supported by null results will always be less incisive than one based on positive ones. It emerges thus that inertial systems, on which basically rely special relativity, are inappropriate to study the effects of their own speed, since local sources and observers having the same speed, the net effect is always null. The way to reveal the effect of speed is by means of rotating systems, which allow breaking symmetry between the source and observer speeds. It appears thus essential to overcome the inability of inertial systems and to appeal instead to rotating ones. These offer also the great advantage of providing an absolute reference embodied in the rotor axis, since on it the peripheral speed is null. So, the effect of motion can be referred to this absolute reference, allowing hence to distinguish between the speed of the source and that of the observer.

Let us hence stress and sum up the great advantages offered by rotary systems in sounding effects due to speed, allowing to analyse separately the influence of the source speed on light emission and of the observer speed on its detection. When the observer is fixed on the rotor axis since its peripheral speed is null any observed effect is exclusively due to the source speed. Oppositely, when the source is fixed on the rotor axis any effect then observed is exclusively due to the observer peripheral speed. This contrasts with the limitations of inertial systems, in which any effect can only be expressed in terms of relative speed, without allowing to discern each speed and to identify and ponder separately the influence of each one. Furthermore, if source and observer are diametrically set, their peripheral speed vector then have always opposite directions, and thus at any infinitesimal instant, the rotary system simulates two vehicles going in opposite directions, at the same speed with respect to a common absolute reference constituted by the rotor axis. Hence the effects of speed do not cancel out but they add, and thus turn out observable. The rotary system presents also the great advantage of ascertaining the effect of speed by turning it on and off. It would thus be of great interest to construct such an ultra-fast rotary system and sound the effect of speed, overcoming the limitations of relaying on relative speeds. For example, on an inertial system interchanging source and observer has no effect, but on a rotary system it does, widening our apprehension of the causality of effects.

On inertial systems since no net speed-induced effect from local sources has been observed, special relativity has handled the proper system as free of speed. However, this standpoint deeply falsifies the factual reality. Actually, the proper inertial system is moving through space even though its motion is not detected, so during the beam time-of-flight the detector has slightly moved away. However, since no aberration is observed this necessarily implies the beam to be pointing forward, reaching so the detector just as if the inertial system had effectively no speed at all nor the beam any forward deflection. In order to solve this pernicious ambiguity, it has been appealed to a rotatory experiment, which brakes the symmetry between the source and detector speed, allowing so to evidence the speed-induced deflection of light rays and to evade the inadequacy of inertial systems due to the systematically null results they supply. This would allow to amend the improper conceptual standpoint extrapolated from null results, and to upgrade it with more acute deductions from non-null results. An experiment able to evidence the speed-induced deflection of light rays appears thus of fundamental welfare.

**Figures**

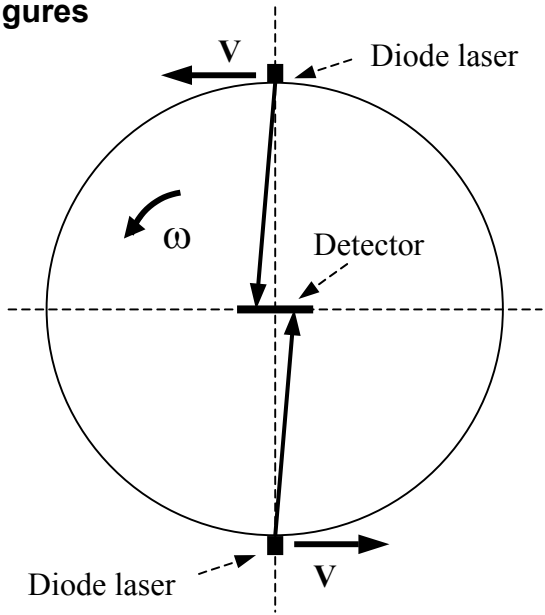

**Fig.1**

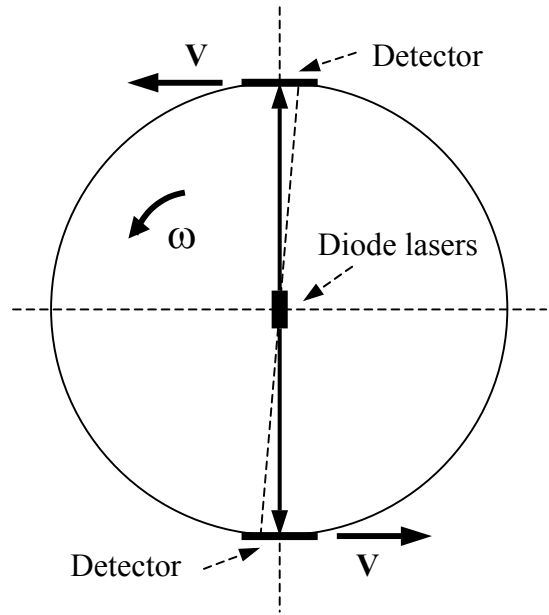

**Fig.2**

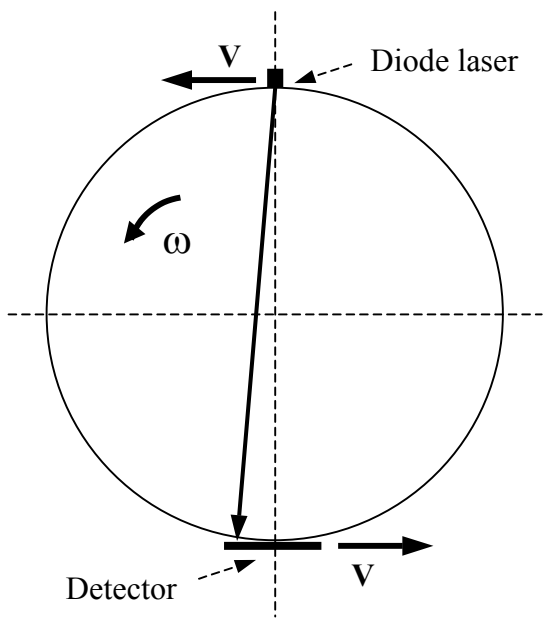

**Fig.3**